\documentclass[sigconf, nonacm]{acmart}

\AtBeginDocument{%
  }

\setcopyright{acmlicensed}
\copyrightyear{2026}
\acmYear{2026}
\acmDOI{XXXXXXX.XXXXXXX}
\acmConference[Conference acronym 'XX]{Make sure to enter the correct
  conference title from your rights confirmation email}{June 03--05,
  2026}{Woodstock, NY}

\acmISBN{978-1-4503-XXXX-X/2018/06}

\usepackage{dirtytalk}
\usepackage{multirow}
\usepackage{booktabs}
\usepackage{tikz}
\usetikzlibrary{arrows.meta, positioning}

\begin{document}

\title{Sociotechnical Challenges of Machine Learning in Healthcare and Social Welfare}

\author{Tyler Reinmund}
\email{tyler.reinmund@jesus.ox.ac.uk}
\orcid{0000-0002-9249-7688}
\affiliation{%
  \institution{University of Oxford}
  \city{Oxford}
  \country{UK}
}
\author{Lars Kunze}
\email{Lars.Kunze@uwe.ac.uk}
\orcid{https://orcid.org/0000-0001-5302-1938}
\affiliation{%
  \institution{University of the West of England}
  \city{Bristol}
  \country{UK}
}
\author{Marina Jirotka}
\email{marina.jirotka@cs.ox.ac.uk}
\orcid{https://orcid.org/0000-0002-6088-3955}
\affiliation{%
  \institution{University of Oxford}
  \city{Oxford}
  \country{UK}
}

\renewcommand{\shortauthors}{Reinmund et al.}

\begin{abstract}
Sociotechnical challenges of machine learning in healthcare and social welfare are mismatches between how a machine learning tool functions and the structure of care practices. While prior research has documented many such issues, existing accounts often attribute them either to designers’ limited social understanding or to inherent technical constraints, offering limited support for systematic description and comparison across settings. In this paper, we present a framework for conceptualizing sociotechnical challenges of machine learning grounded in qualitative fieldwork, a review of longitudinal deployment studies, and co-design workshops with healthcare and social welfare practitioners. The framework comprises (1) a categorization of eleven sociotechnical challenges organized along an ML-enabled care pathway, and (2) a process-oriented account of the conditions through which these challenges emerge across design and use. By providing a parsimonious vocabulary and an explanatory lens focused on practice, this work supports more precise analysis of how machine learning tools function and malfunction within real-world care delivery.
\end{abstract}



\maketitle

\section{Introduction}
\label{sec:Introduction}

Over the past decade, the desire to apply machine learning-enabled tools (ML tools) to a range of problems affecting the United Kingdom's healthcare and social welfare sectors has reached a crescendo \cite{topol2019digital}. Yet, despite the increasing interest, adoption of these technologies remains in early stages \cite{nao2024ai}. Apart from the well-documented technical, organizational, and ethical issues that undermine the effectiveness, quality, and safety of ML-enabled care services \cite{blake2021nhsx, nao2024ai, topol2019digital}, there is growing recognition that a subset of related complications arise through a combination of social and technical phenomena, what scholars have taken to calling \textbf{sociotechnical challenges} \cite{zajac2023clinician}. We define sociotechnical challenges as those that occur as machine learning tools are integrated and repeatedly used in the working practices of care professionals, arising from dissonances between an ML tool -- as it is developed and used -- and the norms, interpretive schemes, and facilities that structure professional practice. We focus on challenges observable during use, rather than during model training or abstract ethical risks.

In this paper, we aim to extend existing discussions on sociotechnical challenges by presenting a categorization and process model. The categorization delineates 11 sociotechnical challenges of machine learning that have been observed in healthcare and social welfare; each challenge is connected to their practical consequences for performance, reliance, and work practices. The process model depicts three paths through which sociotechnical challenges may emerge and the conditions influencing their occurrence. Together, this framework aims to support the description and explanation of the sociotechnical challenges of machine learning by offering a shared vocabulary for analyzing ML deployment beyond technical failure.

\section{Background}
\label{sec:Background}

Across assorted writings in Human-Computer Interaction (HCI), Computer-Supported Cooperative Work (CSCW), and Fairness, Accountability, and Transparency in Machine Learning (FAccT ML), scholars have generated several conceptual models for explaining the emergence of sociotechnical challenges. In their writings, authors propound two variants: the first argues that sociotechnical challenges are the consequence of designers' fallible mental models, while the second suggests that fundamental limitations in technology are to blame.\footnote{In the literature on sociotechnical challenges, authors tend to use the term \say{designer} to refer to the collective of technical personnel responsible for the design and development of a technology. Distinctions between roles such as user researchers, engineers, programmers, interface designers, and system architects are consequently abstracted away. In this paper, we follow the literature's convention with usage of the term designer but explicitly bring attention to the diversity of occupational roles subsumed under this general category.} In this section, we critically review each account and suggest paths forward for a more nuanced theoretical understanding.

The version gaining the most traction is one which we refer to as the \textbf{problem of social understanding} \cite{selbst2019fairness, green2020algorithmic}. In this account, designers draw on mental models and implicit theories of the world to create artifacts that are ultimately incompatible with the realities of their users, thus leading to sociotechnical challenges. These cognitive resources -- imparted on designers through computer science education -- are taken to be deficient: fallible and simplistic abstractions of a complex and nuanced reality. For example, designers may prioritize the performance of the technical system over other organizational goals, assume that complex social phenomena can be unproblematically rendered into technical representations, and fail to consider how technologies induce change within social structures \cite{selbst2019fairness, green2020algorithmic}.

Although the second model has not yet been introduced to the discussion on sociotechnical challenges in machine learning, it is worth mentioning here as it plays a crucial role in informing how HCI and CSCW scholars conceive of the relation between the worlds of social actors and their technical artifacts. The \textbf{social-technical gap}, popularized by \citet{ackerman2000intellectual}, provides a more generous reading of technologists as compared to the problem of social understanding; in contrast, designers have an appreciation of the complexity of the social world, but run up against the fundamental constraints of technology. The brittleness and ridigity of technologies are incompatible with the flexibility and ambiguity of human behavior. Understanding this intractable gap between \say{what we know we must support socially and what we can support technically} is what \citet[p. 180]{ackerman2000intellectual} proposes to be the central challenge of CSCW.

The writings on sociotechnical issues in machine learning provide a productive scaffolding for the problem of sociotechnical challenges. Yet, existing models suffer from incompleteness. The problem of social understanding and the social-technical gap both usefully reveal how unsophisticated conceptualizations of people or intractable constraints of technology lead to socially incompatible technologies, respectively. Nonetheless, each model is unable to account for the process articulated by the other, and both so far have explained sociotechnical challenges as a problem due to design, neglecting how the actions of users may affect their emergence. 

\section{Methods}

We employed a four-phase study to develop a theoretical model of sociotechnical challenges of machine learning in healthcare and social welfare. After establishing a theoretical foundation to guide our data collection and analysis, we conducted fieldwork on the implementation of an ML tool within a preventive care program; we identified several sociotechnical challenges and the conditions which facilitate their emergence. These concepts were then iteratively refined through engagement with similar longitudinal ML deployment studies and co-design workshops.

\subsection{Phase 1: Theoretical Foundation}

To extend current conceptualizations of sociotechnical challenges, we draw on Technologies-in-Practice: a theory for studying the relationships between technology and organizations \cite{orlikowski2000using}. This theory is particularly well-suited for articulating sociotechnical challenges in machine learning because it accounts for the processes of both technological design and use, and does not grant priority to either people or artifacts \emph{a priori} when explaining the role of technology in organizational change.

Technologies-in-Practice posits that people draw on and enact structures during their routine interactions with technology. This builds on the \say{duality of structure} \cite{giddens1979central, giddens1984constitution}: structure is not merely a constraint on human activity, but is also created by it. On one side of this duality, people use structures to perform social activity; on the other, they reproduce those structures through their activity.

Drawing from Giddens, Orlikowski describes three modalities of structure that mediate people's interactions with technology:
\begin{itemize}
    \item \textbf{Facilities:} The material properties of the technology, including those inscribed by designers and those added by users \cite[p. 410]{orlikowski2000using}. While users can devise workarounds or modify properties, the realm of possible adaptations is bound by the technology's physical constraints.
    \item \textbf{Interpretive schemes:} The cognitive characteristics of people, including the \say{mental models or frames of reference that individuals have about the world, their organization, work, technology, and so on} \cite[p. 364]{orlikowski1992learning}.
    \item \textbf{Norms:} The \say{organizational conventions or rules governing legitimate or `appropriate' conduct} \cite[p. 405]{orlikowski1992duality}, such as reward systems, protocols, and etiquette.
\end{itemize}

\subsection{Phase 2: Fieldwork}

Between December 2022 and May 2023, our team engaged in fieldwork with a social welfare organization in England deploying an ML tool as part of a fall prevention program for older adults. Over this period, the lead author conducted over 100 hours of participant observation, held 53 formal and informal semi-structured interviews with data scientists, project managers, consultants, and service administrators, and reviewed internal project documentation and artifacts. We discuss the field site and data collection in greater detail in earlier work, as well as describing the program implementation process and operational obstacles \cite{reinmund2024transitioning, salvini2023human}.

\subsection{Phase 3: Review of Longitudinal Studies}

We conducted a qualitative review of 22 peer-reviewed longitudinal studies on the implementation and use of machine learning tools within care pathways. Each study can be classified into one of three types:

\begin{enumerate}
    \item \textbf{Case reports:} These studies present findings derived from applied projects on the implementation of machine learning tools in healthcare and social welfare pathways \cite{engstrom2022operationalizing, gonccalves2020implementation, murphree2021improving, petitgand2020investigating, pou2022compute, romero2020implementation, ruamviboonsuk2022real, salvini2023human, sendak2020human, sendak2020real, reinmund2024transitioning}.
    \item \textbf{User experience evaluations:} These studies are aimed at developing bounded generalizations regarding experiences of use with a specific machine learning tool through single- and multi-sited evaluations \cite{beede2020evaluation, cheng2022child, deartega2020case, kawakami2022improving, sun2023data, wang2021brilliant}.
    \item \textbf{Ethnographies:} These studies report ethnographic observations on the implementation and use of machine learning tools, predominantly in a health care context, and the resultant changes in communication, collaboration, and organization among social actors \cite{elish2018stakes, elish2020repairing, ismail2023public, lebovitz2022engage, maiers2017analytics}.
\end{enumerate}

\subsection{Phase 4: Co-Design Workshops}

To validate, amend, and extend the model of sociotechnical challenges, we conducted 5 co-design workshops with 15 healthcare and social welfare practitioners. The objective of the workshop was to collaboratively define a dimension across which to distinguish the sociotechnical challenges.

The co-design workshop protocol included both asynchronous and synchronous components. Prior to the workshop, participants individually completed a closed, online sorting activity \cite{spencer2009card} to familiarize themselves with the workshop topics. This activity was conducted through an online questionnaire in which participants categorized each challenge in response to the following question: \emph{when during the process of interacting with a ML tool is this challenge most likely to arise?} Participants then joined an online, synchronous workshop in which they discussed the sociotechnical challenges in greater detail and resolved disagreements over their categorization. This dimension structures the presentation of the 11 sociotechnical challenges discussed in the next section.

\section{Sociotechnical Challenges Across the ML-enabled Care Pathway}
\label{sec:Challenges}

Grounded in our fieldwork and refined through the subsequent literature review and workshops, we identified 11 sociotechnical challenges positioned along various stages of the ML-enabled care pathway. While we distinguish these challenges analytically, in the empirical reality of care work they often overlap. The ML-enabled care pathway refers to the sequence of activities performed as practitioners use ML tools within care delivery, assuming the tool is deployed and making inferences. The sociotechnical challenges are summarized in Table~\ref{tab:sociotechnical-challenges}.

\begin{table*}[htbp]
\renewcommand{\arraystretch}{1.15}
\centering
\small
\begin{tabular}{@{}p{1.8cm}p{3.6cm}p{8.2cm}@{}}
\toprule
\textbf{Stage} & \textbf{Sociotechnical Challenge} & \textbf{Description} \\
\midrule
\multirow{3}{*}{Gather Data}
& Data Inconsistencies 
& Variations in data formats, units, or collection practices reduce the reliability of model inputs, often degrading predictive performance. \\

& Poor Environmental Conditions 
& Physical workplace conditions (e.g., lighting, temperature) compromise data quality, reducing model accuracy and limiting the effectiveness of resulting care decisions. \\

& Data Restrictions 
& Ethical, legal, or organizational limits on data access constrain what can be modeled, reducing predictive performance and limiting the effectiveness of resulting care decisions. \\
\midrule
\multirow{3}{*}{Interpret Output}
& Hidden Predictors 
& Practitioners rely on contextual cues absent from the model, which can reduce reliance when omissions are recognized or outputs conflict with judgment. \\

& Mismatched Objectives 
& Differences between practitioners’ decision goals and the model’s prediction target lead to selective or inconsistent reliance. \\

& Decision Point Disconnect 
& ML outputs fail to reach intended users at moments of decision-making, resulting in reduced usage. \\
\midrule
\multirow{2}{*}{Initiate Action}
& Insufficient Causal Information 
& Practitioners attempt to use an ML tool to answer cause-and-effect questions for which it was not designed, reducing decision effectiveness and use. \\

& Value Conflicts 
& Tool use reveals conflicting interpretations of organizational values among stakeholders, prompting changes in norms, roles, or tool design. \\
\midrule
\multirow{3}{*}{Multi-Stage}
& Disrupted Workflows 
& ML tools reconfigure communication and coordination practices, reducing use and reshaping power relations across care teams. \\

& Limited Understanding 
& Incomplete understanding of how outputs are produced leads to reduced reliance and performance, while also prompting informal, ad hoc learning. \\

& Additional Labor 
& Tool use requires additional, often unexpected forms of data and coordination work, decreasing efficiency and usage while encouraging workarounds and conditional reliance. \\
\bottomrule
\end{tabular}
\caption{Sociotechnical challenges identified across stages of the ML-enabled care pathway. While analytically distinct, challenges often overlap in practice.}
\label{tab:sociotechnical-challenges}
\end{table*}

\subsection{Gather Data}

The pathway begins with gathering data, where domain-relevant information is collected (automatically or manually) and transferred into the tool. We identified three challenges that tend to manifest during this stage.

\textbf{Data Inconsistencies} occur when input data deviate from expected structures, types, or units. For example, a case report on a patient deterioration warning system noted how the hospital changed the sensitivity of its troponin assay during implementation, leading to discrepancies in feature scaling \cite{pou2022compute}. Such inconsistencies reduce the reliability of model inputs and often degrade predictive performance.

\textbf{Poor Environmental Conditions} arise when the physical workplace environment degrades the quality of data processed by the tool. In an evaluation of a diabetic retinopathy screening tool, \citet{beede2020evaluation} observed how images captured in non-darkened environments suffered from decreased quality. These conditions compromise data quality, reducing model accuracy and limiting the effectiveness of subsequent care decisions.

\textbf{Data Restrictions} imply that access to data required for predictions is limited or ethically fraught. For instance, a public health initiative in India faced the dilemma of identifying individuals excluded from care due to caste without collecting that information explicitly, which would leave them vulnerable to data misuse \cite{ismail2023public}. Such restrictions constrain what can be modeled, reducing predictive performance and limiting the effectiveness of resulting care decisions.

\subsection{Interpret Output}

The input data are analyzed by the tool to produce an output (e.g., risk scores, recommendations) which is then interpreted by professionals. Three challenges are prominent here.

\textbf{Hidden Predictors} describe situations where end-users rely on contextual information unobserved by the ML tool, such as visual cues \cite{wang2021brilliant} or performance metrics \cite{yang2019unremarkable}. When users recognize that relevant factors are omitted, their reliance on the tool decreases \cite{deartega2020case, cheng2022child}.

\textbf{Mismatched Objectives} occur when end-users' decision-making objectives differ from the variable predicted by the tool. For example, case workers in child welfare may prioritize short-term immediate risk, whereas the ML tool may prioritize long-term risk of re-referral \cite{kawakami2022improving, cheng2022child}. These differences lead to selective or inconsistent reliance on the tool’s outputs.

Finally, \textbf{Decision Point Disconnect} occurs when the intended user does not come in contact with the tool when making decisions. \citet{petitgand2020investigating} reported how clinicians rarely received generated medical histories because overworked nurses lacked the capacity to print and deliver the reports. As a result, tool usage is reduced despite technical availability.

\subsection{Initiate Action}

In this stage, an action (intervention or monitoring) is initiated based on the output. Two challenges manifest here.

\textbf{Insufficient Causal Information} suggests end-users attempt to use a tool to answer causal questions for which it was not designed. During our fieldwork, we observed stakeholders appropriating feature importance measurements to anticipate recommended care services, only to find the features were not \say{actionable} or amenable to intervention. This misalignment reduces decision effectiveness and discourages continued use.

\textbf{Value Conflicts} occur when use of the tool reveals conflicting interpretations of organizational values. During our workshops, a participant recounted how the evaluation of an ML tool for ophthalmologists surfaced conflicts over a clinician's right to dissenting professional judgment. Such conflicts prompt changes in norms, roles, or tool design rather than straightforward adoption.

\subsection{Multi-Stage}

The final category refers to challenges manifesting across the pathway.

\textbf{Disrupted Workflows} implies that use of the tool challenges pre-existing lines of communication. \citet{elish2020repairing} explained how an ML tool affected diagnostic processes by obviating the need for face-to-face interaction during sepsis diagnosis. These disruptions can reduce use and reshape power relations across care teams.

\textbf{Limited Understanding} occurs when end-users struggle to understand how the tool functions. \citet{lebovitz2022engage} illustrated how a lack of interface information led radiologists to struggle with predictions diverging from their judgment. Limited understanding reduces reliance and performance while also prompting informal, ad hoc learning.

\textbf{Additional Labor} involves the creation of previously unperformed work. \citet{sun2023data} described the intensive \say{data work} required of care workers to collect and input data for an ML tool. This additional labor decreases efficiency and usage while encouraging workarounds and conditional reliance.

\section{Conditions Influencing the Emergence of Sociotechnical Challenges}

Sociotechnical challenges are not static outcomes; they emerge through dynamic interactions between people, technology, and practice. Drawing on our fieldwork, synthesis of the literature, and co-design workshops, we propose a multidimensional process model of sociotechnical challenges. This model functions as interpretive scaffolding to help researchers and practitioners identify the conditions under which the challenges described in Section~\ref{sec:Challenges} are likely to arise. We explicate three specific processes: determining design, grappling with constraints, and deviating from scripts.

\subsection{Determining Design}

The first process refers to the social and political negotiations where stakeholder values and organizational pressures are translated into the material properties of an ML tool. These properties are not inevitable; they are conditioned by the technical competence of designers, the distribution of decision-making authority, and how work tasks are formalized. When authority is concentrated in stakeholders distant from the point of care, the resultant design often fails to account for the nuances of clinical practice. For instance, \textbf{Mismatched Objectives} often arise through Determining Design, where predictive targets are fixed early based on managerial priorities, such as long-term cost savings, rather than the immediate decision-making needs of frontline staff.

\subsection{Grappling with Constraints}

Building on the social-technical gap \cite{ackerman2000intellectual}, this process highlights instances where the rigid physical and technical limitations of an artifact clash with the fluidity of care work. Designers must contend with the validity of statistical representations, the stability of the physical operating environment, and the interoperability of underlying data systems. The limitations of representing complex social phenomena as quantitative measures often lead to friction during use. For example, \textbf{Insufficient Causal Information} frequently manifests as users grapple with representational constraints, attempting to derive actionable interventions from features that were selected solely for their statistical predictive power rather than their clinical relevance.

\subsection{Deviating from Scripts}

Once an ML tool is deployed, users integrate the technology into their practice through improvisation rather than strict adherence to the scripts envisioned by designers. Through processes of appropriation, peer coordination, and enrollment, users may modify their workflows or the tool itself to fit local circumstances. However, this improvisation can lead to dissonant relationships between the tool and established norms. \textbf{Hidden Predictors}, for example, emerge as users deviate from the intended reliance on the tool's output, choosing instead to prioritize contextual knowledge, such as visual cues or patient history, that they realize the model has failed to capture.

\section{Conclusion}

In this paper, we introduced a framework for the sociotechnical challenges of machine learning in healthcare and social welfare. This framework includes a classification of 11 observed challenges and a process model characterizing the conditions influencing their occurrence. The objective is to provide a parsimonious set of concepts to enable the description of complications affecting the integration of ML tools into care pathways.

We note that the process model presented here is provisional. It represents a high-level synthesis of a broader theoretical elaboration detailed in forthcoming work. By offering this initial scaffolding, we aim to move the discourse toward a concrete understanding of how ML tools function and malfunction within the messy reality of professional care practice.



\bibliographystyle{ACM-Reference-Format}
\bibliography{biblio}

\end{document}